\documentclass[useAMS,usenatbib]{mn2e}
\usepackage{graphicx}
\usepackage[]{natbib}

\def \hcm {\hbox {\ifmmode $ atom cm$^{-2}\else atom cm$^{-2}$\fi}}

\def \apj {ApJ}
\def \apjl {ApJL}
\def \aap {A\&A}

\def \mnras {MNRAS}

  \title{A Swift view on IGR~J19149+1036.}
\author[G.\ Cusumano et al.]{G.\ Cusumano$^{1}$,  A.\ Segreto $^{1}$ , V.\ La Parola$^{1}$, N.\ Masetti
$^{2}$, A.\ D'A\`i, G.\ Tagliaferri$^{3}$ \\
$^{1}$INAF - Istituto di Astrofisica Spaziale e Fisica Cosmica, Via U.\ La Malfa 153, I-90146 Palermo, Italy\\
$^{2}$INAF - Istituto di Astrofisica Spaziale e Fisica Cosmica di Bologna, via Gobetti 101, 40129, Bologna, Italy\\
$^{3}$  INAF - Brera Astronomical Observatory, via Bianchi 46, 23807, Merate (LC), Italy
\\
}

\begin{document}
\date{}

\pagerange{\pageref{firstpage}--\pageref{lastpage}} \pubyear{2014}

\maketitle
            
\label{firstpage}
\begin{abstract}

IGR~J19149+1036 is a high mass X-ray binary detected by INTEGRAL 
in 2011 in the hard X-ray domain.  
We have analyzed the BAT survey data of the first 103 months of the Swift mission detecting this
source at a significance level of $\sim30$ standard deviations. The timing analysis on the long term
BAT light curve reveals the presence of a strong sinusoidal intensity modulation of $22.25\pm 0.05$ d, that we
interpret as the orbital period of this binary system.

A broad band (0.3-150 keV) spectral analysis was performed combining the BAT spectrum and the
XRT spectra from the pointed follow up observations. The spectrum is adequately modeled with
an absorbed power law with a high energy cutoff  at $\sim24$ keV and an absorption 
cyclotron feature at $\sim31$ keV. Correcting 
for the gravitational redshift, the inferred  magnetic field at the neutron star surface is 
$B_{\rm surf} \sim 3.6 \times 10^{12}$ gauss.

\end{abstract}

\begin{keywords}
X-rays: binaries -- X-rays: individual: IGR~J19149+1036

\noindent
Facility: {\it Swift}

\end{keywords}

\section{Introduction}\label{intro} 

Since 2004 November, the Burst Alert Telescope (BAT, \citealp{bat}) on board the 
Swift observatory  \citep{swift} has been scanning the 
entire sky in the hard X-ray domain (15--150 keV) recording 
timing and spectral information for any detected source (more than 1700 sources after 100
months of survey\footnote{http://bat.ifc.inaf.it}). Thanks to its wide field of view 
and to the Swift pointing strategy, the BAT observes daily $\sim 90\%$ of the sky, and
therefore it is specially fit for source variability studies.
BAT has proved to be a valuable tool to detect transient phenomena 
from known and unknown sources \citep{krimm13}
and, by combining the entire data span, to reveal long 
periodicities of Galactic high mass X-ray binaries (HMXB, e.g. \citealp{corbet1, corbet2, 
corbet3, corbet4, corbet5, corbet6, corbet7, cusumano10, cusumano13a, cusumano13b, laparola10,
laparola13, segreto13a,segreto13b, dai11}).

In this paper we present a comprehensive temporal and spectral analysis of the Swift data
collected on IGR~J19149+1036. 
This source was detected by IBIS/ISGRI at a  position consistent 
with the Einstein source 2E~1912.5+1031  \citep{pavan11}, although a firm association was not
possible due to large systematic uncertainties related to the presence of GRS 1915+105 at a 
distance lower than $\sim 20'$. The source was 
also detected by JEM-X with a 3-10 keV X-ray flux of $\rm 7\times 10^{-12} erg~s^{-1}cm^{-2}$.
The field around IGR~J19149+1036 was observed by Swift/XRT several times.
The analysis of the XRT data collected on 2010 Dec. 5 and 
2011 Feb. 17, reported in \citet{bozzo11}, allowed to estimate a refined source position at 
RA = 19h14m56.73s and Dec = +10deg 36' 38.11'' (J2000), with an associated uncertainty 
of 3.7 arcsec, that confirmed the association with 2E~1912.5+1031 suggested 
by \citet{pavan11}. The XRT spectrum was modeled by an absorbed 
power-law with a column density N$_{\textrm H}=\rm 6.5 \times 10^{22} cm^{-2}$ and a
photon index $\Gamma=1.8$, with an observed 1-10 keV X-ray flux of
$\rm 1.8\times 10^{-11} erg~s^{-1}cm^{-2}$. 2MASS~J19145680+1036387 (J=14.34,
H=12.41, K=11.53), within the XRT error circle, is the most likely  counterpart 
to IGR~J19149+1036. 

This paper is organized as follows. Section 2 describes the BAT and XRT data reduction. 
Section 3 reports on the timing analysis. Section 4 describes the broad
band spectral analysis. In Section  5 we briefly discuss our results.

\begin{figure}
\begin{center}
\centerline{\hspace{.5truecm}\includegraphics[width=7.5cm,angle=0]{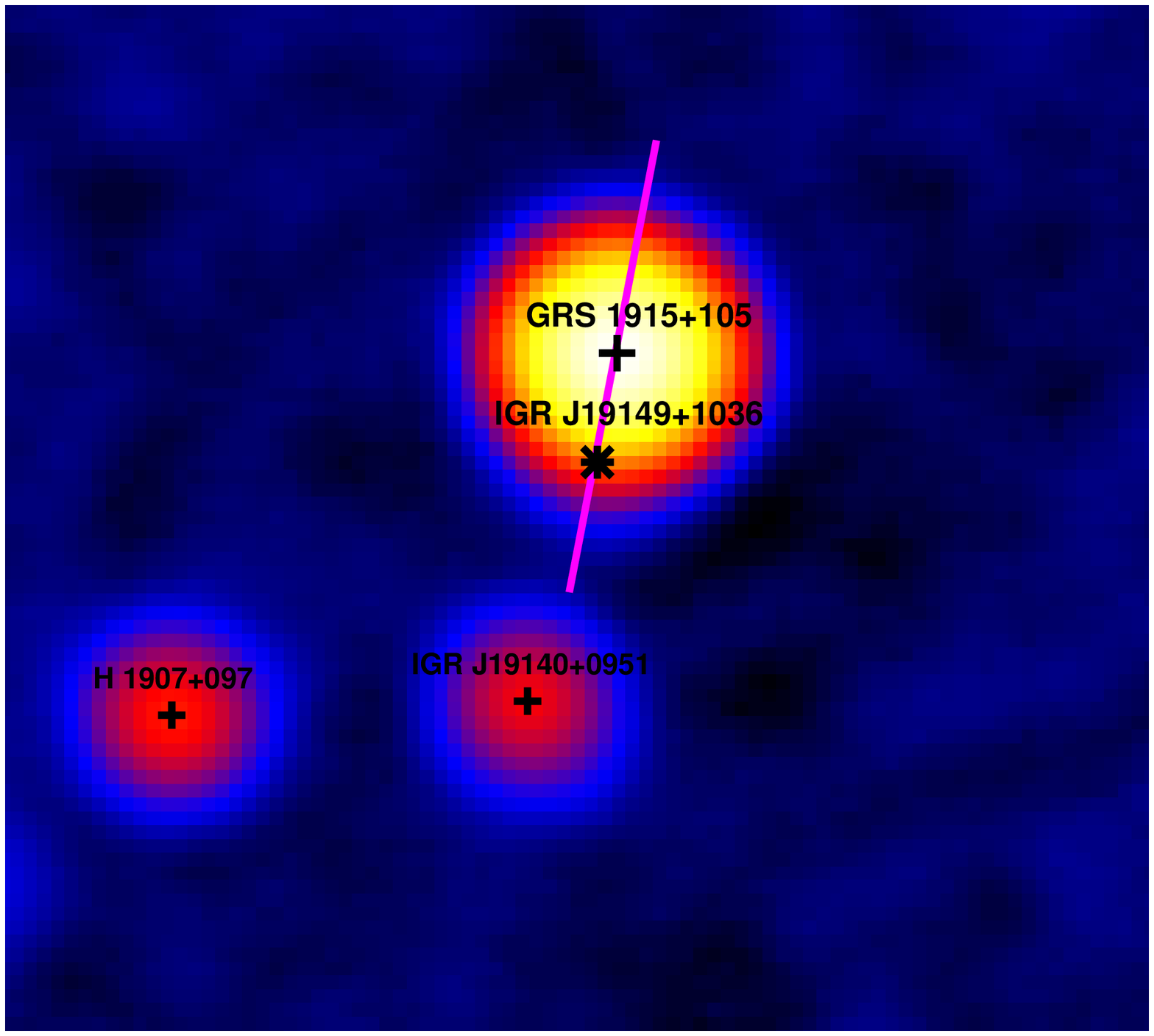}}
\centerline{\includegraphics[width=8.5cm]{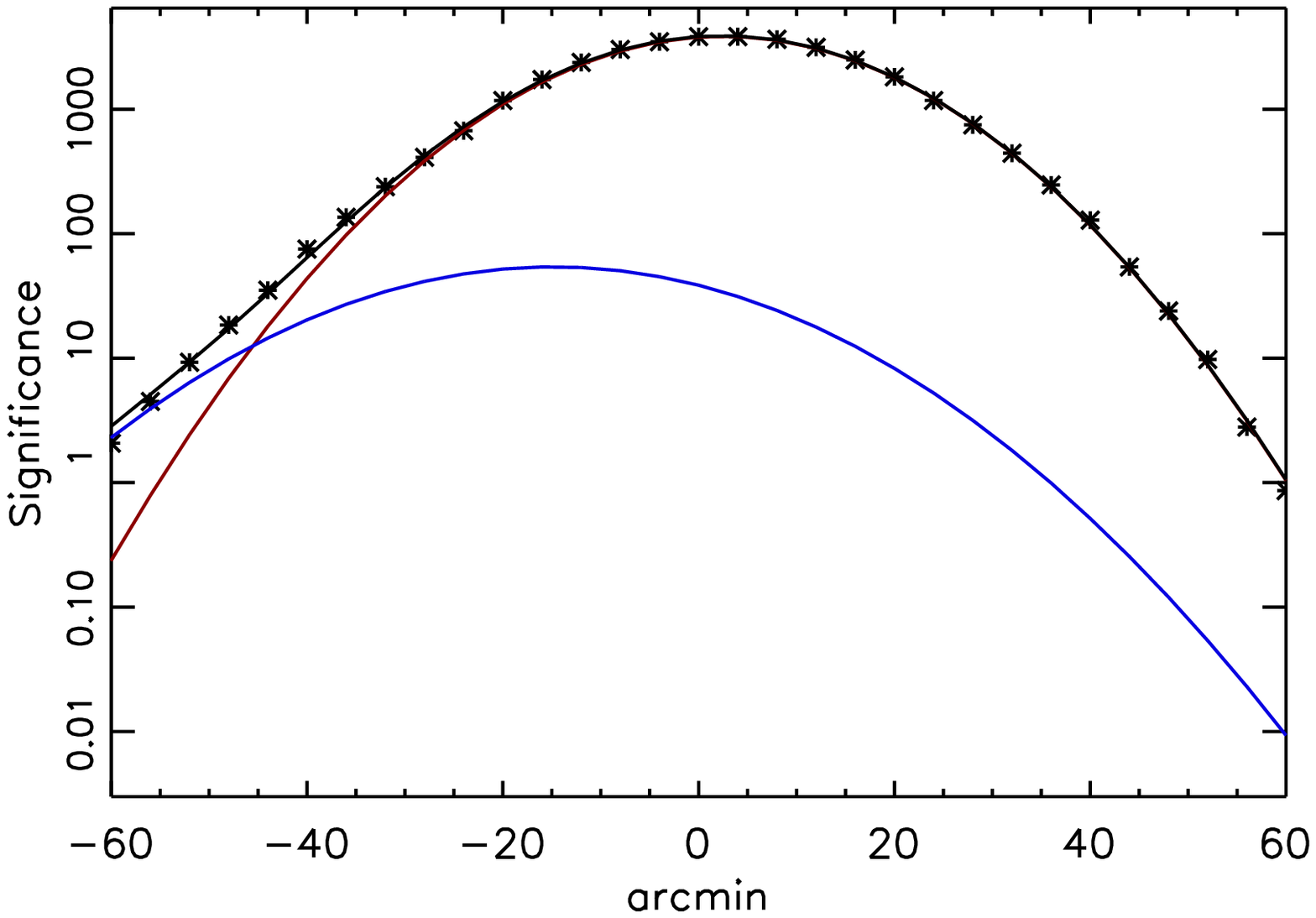}}
\centerline{\hspace{.5truecm}\includegraphics[width=7.5cm,angle=0]{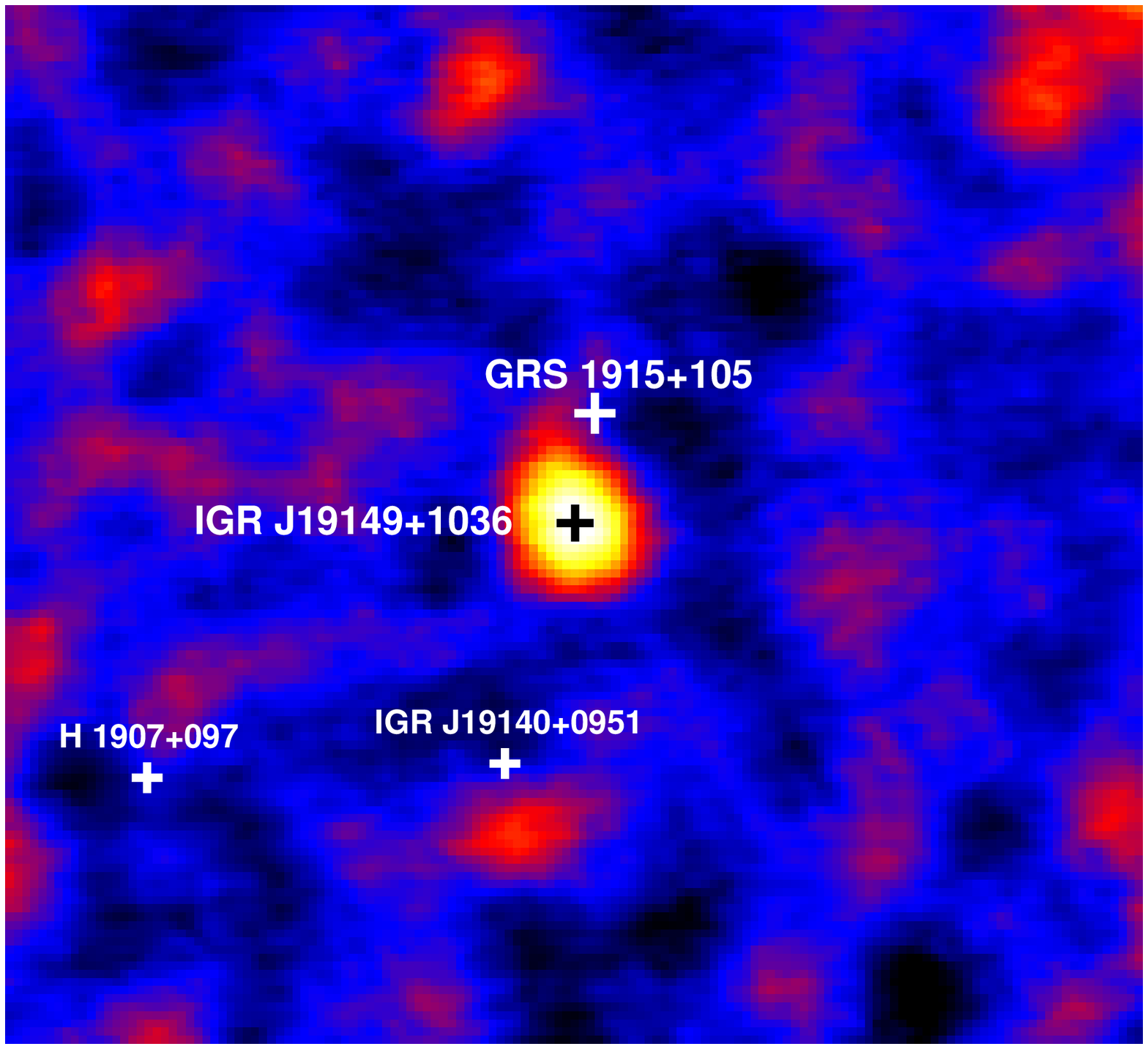}}
\caption[IGR~J19149+1036 sky maps]{ 
{\bf Top panel}:  
15-150 keV BAT sky map around IGR~J19149+1036
The crosses  mark the position of the sources detected in the field; the star
marks the XRT position of IGR~J19149+1036. 
{\bf Middle panel}: 
Values of the significance extracted along the line that connects the 
optical position of the two surces (magenta
line in top panel). The plot shows the data (crosses) and the best-fitting
model (black line, the sum of two Gaussian profiles).
The higher peak (red line) corresponds to GRS~1915+105 while the lower peak (blue line)
corresponds to IGR~J19149+1036.
{\bf Bottom panel}: 15-150 keV significance "residual map".
                }
                \label{map} 
        \end{center}
        \end{figure}

\section{Observations and data reduction}\label{data}

The results reported in this paper on IGR~J19149+1036 derive from the analysis of the
BAT survey data relevant to the  first 103 months of the Swift
mission. The data were retrieved from the Swift public
archive\footnote{http://heasarc.gsfc.nasa.gov/cgi-bin/W3Browse/swift.pl}
in the form of detector plane histograms (DPH):
three-dimensional arrays (two spatial dimensions, one spectral
dimension) that collect count-rate data in 5-min time bins for
80 energy channels. We use a dedicated software \citep{segreto10} 
to process these data, producing all-sky maps in several energy intervals between 15 and 150 keV,
detecting sources on these maps and
extracting, for each source, standard products such as light curves and spectra.
Figure~\ref{map}  (top panel) shows the 15-150 keV significance sky map
(exposure time 39.4 Ms) in the direction of IGR J19149+1036. The source is close
($\sim$20 arcmin) to the much brighter object GRS 1915+105: the positions of these two 
sources are marked with  black crosses. In this map IGR J19149+1036 is not immediately detected
 because it is embedded in the PSF signal of the close brighter source. 
However, if we extract the values of the significance along the line that connects the 
optical position of the two surces (magenta line in top panel of Figure~\ref{map}) we observe an
asymmetric profile  (Figure~\ref{map}, central panel) that suggests the presence of a 
faint source (IGR J19149+1036) hidden by the PSF of the brighter one (GRS 1915+105).
Indeed, the fit of this profile with a single Gaussian shows
systematic residuals; conversely, the profile is  well fitted with a 
double Gaussian. The best fit peaks of the two Gaussians 
correspond to the position of the two sources.
In order to make IGR J19149+1036 emerge in the sky map, we perform on each
DPH a subtraction of the expected illumination pattern from all the  
detected sources; from the resulting "residual shadowgrams"
we generate new survey "residual maps" where any previously undetected
source can emerge. IGR~J19149+1036 was detected in the 15--150 keV all-sky "residual map"
(Fig. 1, bottom panel) with a signal to noise ratio of 28.6 standard deviations and of 31.7 standard 
deviations in the 15--60 keV all-sky map, where its signal-to-noise 
is maximized. 
The 15--60 keV energy band was used to extract 
the light curve with the maximum resolution ($\sim300$ s) allowed by the 
Swift-BAT survey data.  For each time bin the contribution from GRS 1915+105 is subtracted. 
The time tag of each bin, defined as the central time of the bin interval, was
corrected to the solar system barycenter (SSB) by using the task {\sc earth2sun}. 
The background subtracted spectrum averaged over the entire survey period was 
extracted in 14 energy channels and analyzed using the BAT redistribution 
matrix available in the Swift calibration 
database\footnote{http://swift.gsfc.nasa.gov/docs/heasarc/caldb/swift/}.

Swift-XRT \citep{xrt} observed the field around IGR~J19149+1036  5 times after the discovery of the source
by Integral.  The source was always observed 
in Photon Counting (PC) mode \citep{hill04}
for a total exposure of $\sim$3800 s. 
The details on the five Swift-XRT observations are reported in Table~\ref{log}.
The XRT data were processed with standard procedures ({\sc xrtpipeline} 
v.0.12.8) using the {\sc ftools} in the {\sc heasoft} package (v 6.15) and the 
products were extracted adopting a grade filtering of 0-12. The source events for timing and
spectral analysis were extracted from a circular region of 20 pixel radius (1 pixel = 2.36'') 
centered on the source position as determined with {\sc xrtcentroid} 
(RA= 19h14m56.8s and Dec = +10deg 36' 39.1'', J2000), while 
the background was extracted from an annular region 
centered on the source, with an inner radius of 60 pixels and an outer radius of
 90 pixels that avoids  contamination due to the stray--light from GRS~J1915+105 and to the PSF tail
of IGR~J19149+1036. 
All source event arrival times were converted to 
the SSB with the task {\sc barycorr}\footnote{http://http://heasarc.gsfc.nasa.gov/ftools/caldb/help/barycorr.html}.
Ancillary response files were generated 
with {\sc xrtmkarf}\footnote{http://heasarc.gsfc.nasa.gov/ftools/caldb/help/xrtmkarf.html}. 
We also extracted a single source and a single background spectra from the 5 XRT observations
and combined the relevant ancillary files using {\sc addarf}, that weights them by the exposure 
times of the corresponding source spectra. Finally, the spectra were re-binned with a 
minimum of 20 counts per energy channel. This lower limit of counts per bin is enough to 
ensure that the deviation of the observed number of counts from the expected values 
approximates quite well a Gaussian distribution, that is a requirement to apply the  
$\chi^2$ statistics. We used the spectral redistribution matrix v014 and the 
spectral analysis was performed using {\sc xspec} v.12.5.  

Errors are at 90\,\% confidence level for a single parameter, if not stated otherwise.

\begin{table*}
\caption{XRT observations log. The quoted orbital phase refers to the profile shown   
in Figure~\ref{period}.\label{log}} 
\scriptsize
\begin{center}
\begin{tabular}{r l l l l l l} \hline
Obs \# & Obs ID     & $T_{start}$ & $T_{elapsed}$ & Exposure & Rate            & Orb. Phase  \\
       &            &  MJD        &  (s)          & (s)      &  (c/s)              &     \\ \hline \hline
1      &00041130001 & 55535.220   & 29310.9       &  911.5   &$0.23\pm0.02$        &0.47  \\
2      &00041130002 & 55609.162   & 46002.1       &  1261.2  &$0.13\pm0.01$        &0.79  \\ 
3      &00041130003 & 56267.370   & 388.6         &  387.1   &$0.13\pm0.02$        &0.39  \\ 
4      &00041130004 & 56631.147   & 52911.4       &  849.1   &$0.10\pm0.01$        &0.75  \\ 
5      &00067133005 & 56632.625   & 403.6         &  402.0   &$0.10\pm0.02$        &0.81  \\ \hline
\end{tabular}
\end{center}
\end{table*}

\begin{figure}
\begin{center}
\centerline{\includegraphics[width=9cm,angle=0]{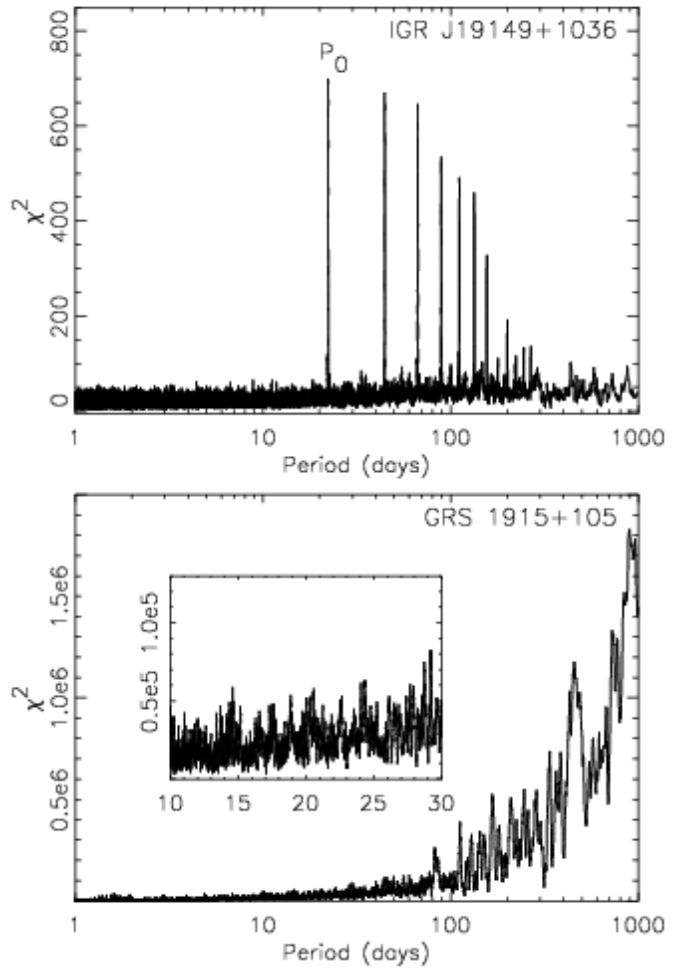}}
\caption[]{{\bf Top panel}: Periodogram of {\it Swift}-BAT (15--60\,keV) data for
IGR~J19149+1036.
{\bf Bottom panel}: Periodogram of {\it Swift}-BAT  (15--60\,keV) data for GRS 1915+105. The inset
shows a close-up view of the periodogram in the 10--30 d interval.                                                              
                \label{period}
}
\end{center}
\end{figure}

\begin{figure}
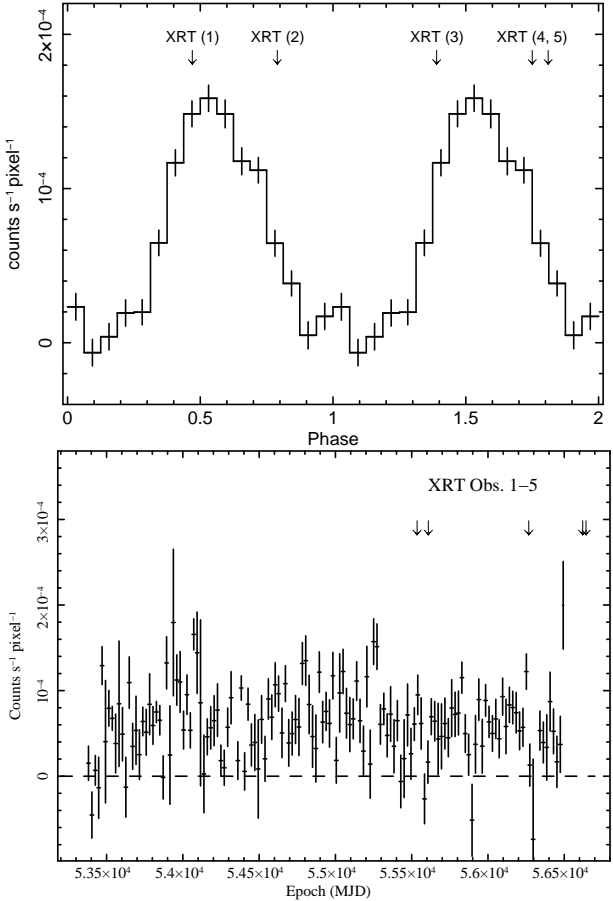

\begin{center}
\centerline{\includegraphics[width=6cm,angle=270]{figura_profilo_pulsato.ps}}
\centerline{\includegraphics[width=6cm,angle=270]{figura_bat.ps}}
\caption[]{
{\bf Top panel}: {\it Swift}-BAT light curve folded at a period $P=22.25\pm0.05$\,d,
with 16 phase bins. The arrow points to the phase corresponding to the time of 
the five XRT observations. 
{\bf Bottom panel}: BAT light curve. The bin length corresponds to a time interval of  
$P_0$\,days.  
                \label{fold}
}
\end{center}
\end{figure}

\section{Timing analysis and results}\label{timing} 

We  produced the periodogram of the long term Swift-BAT 15--60 keV light curve 
applying  a folding technique  \citep{leahy83} which consists in the production of a count rate 
profile at different trial periods by folding the photon arrival times in N phase bins
and evaluating for each resulting profile the $\chi^2$ value with respect to 
the average count rate. A periodic modulation corresponds to a large value of  $\chi^2$.
We searched in the 1--1000 d period range with a period spacing given 
by P$^{2}/(N  \,\Delta$T$_{\rm BAT})$ where P is the trial period, $N=16$  
and $\Delta$T$_{\rm BAT}\sim 272.4$ Ms is the data span length. 
The  average rate in  each profile phase bin was evaluated by weighting the rates  
with  the  inverse  square  of  their statistical error. 
This procedure is mandatory  for data collected by a large field of view coded mask telescope as BAT 
that are characterized by a large spread of statistical errors.
The periodogram (Figure~\ref{period}, top panel)  shows the 
presence of several prominent features. The highest feature is 
at P$_0=22.25\pm0.05$ d ($\chi^2\sim 702$)  
where P$_0$ and its error are the centroid and the standard deviation 
obtained by modeling this feature with a Gaussian function.
Using the method described in \citet{cusumano13a} we find that the probability of finding such a high
$\chi^2$ value by chance is $\sim 10^{-58}$.
The other features clearly visible in  the
periodogram and showing  decreasing $\chi^2$ peaks towards higher trial periods correspond 
to a series of multiples of P$_0$. We verified that the detected periodicity is not present
in the periodogram of GRS 1915+105 as shown in the bottom panel of Fig.~\ref{period}.

The intensity profile (Fig.~\ref{fold}, top panel) folded at  P$_0$ with 
T$_{\rm epoch}=54924.32 $ MJD  is characterized 
by a single symmetric peak with a minimum consistent with zero intensity.
The minimum dip centroid, evaluated by fitting the data 
with a Gaussian function, falls at phase 1.06 $\pm$ 0.01, corresponding
to  MJD $(54925.7\pm 0.2)\pm  n \times $P$_0$. The peak centroid is at phase 0.56$\pm$ 
0.01, corresponding to  MJD $(54936.8\pm 0.2)\pm  n \times $P$_0$ .

Figure~\ref{fold} (bottom panel) shows the 103 months BAT light curve with a time resolution of 
P$_0$ days. The source shows a quite persistent behaviour without any significant
intensity enhancements, which is consistent with the very low level of red noise
observed in the source periodogram shown in figure~\ref{fold}, top panel.

Table~\ref{log} lists the average count rate during each XRT observation and the 
relevant orbital phase evaluated with respect to P$_0$ and
T$_{\rm epoch}$. 
We performed a timing analysis on the XRT data searching for the presence of
a periodic modulation tied to the compact object rotation.
In order to avoid systematics caused by the read-out time in PC mode (characterized by
a time resolution bin of $\delta T_{XRT}$=2.5073 s), the arrival times of the events in 
PC mode were randomized within $\delta T_{XRT}$. 
Moreover, XRT observations are fragmented into snapshots  of different duration and
time separation that may introduce spurious features in the timing analysis. 
To avoid these systematics we performed a folding analysis on each snapshot with 
an exposure time higher than 500 s and with a statistical content larger than 100 counts, 
searching in a period range [$\delta T_{XRT}$~:~100] s. 
The periodograms obtained from snapshots belonging to the same observation
were also summed together. We did not found significant features 
in any of the resulting periodograms.

\section{Spectral Analysis} \label{spectral}
Because the XRT and BAT data are collected at different times and with different exposures,
we verified
that no significant spectral variability is present both among the 
five XRT observations and during the BAT monitoring. To this aim we produced  
hardness ratios  of the source events in the energy bands 4--10 keV / 0.5--4 keV 
from each XRT observation and in the energy bands 35--150 keV / 15--35 keV for 
the BAT data. In both cases, no significant variability is observed.

\begin{figure}
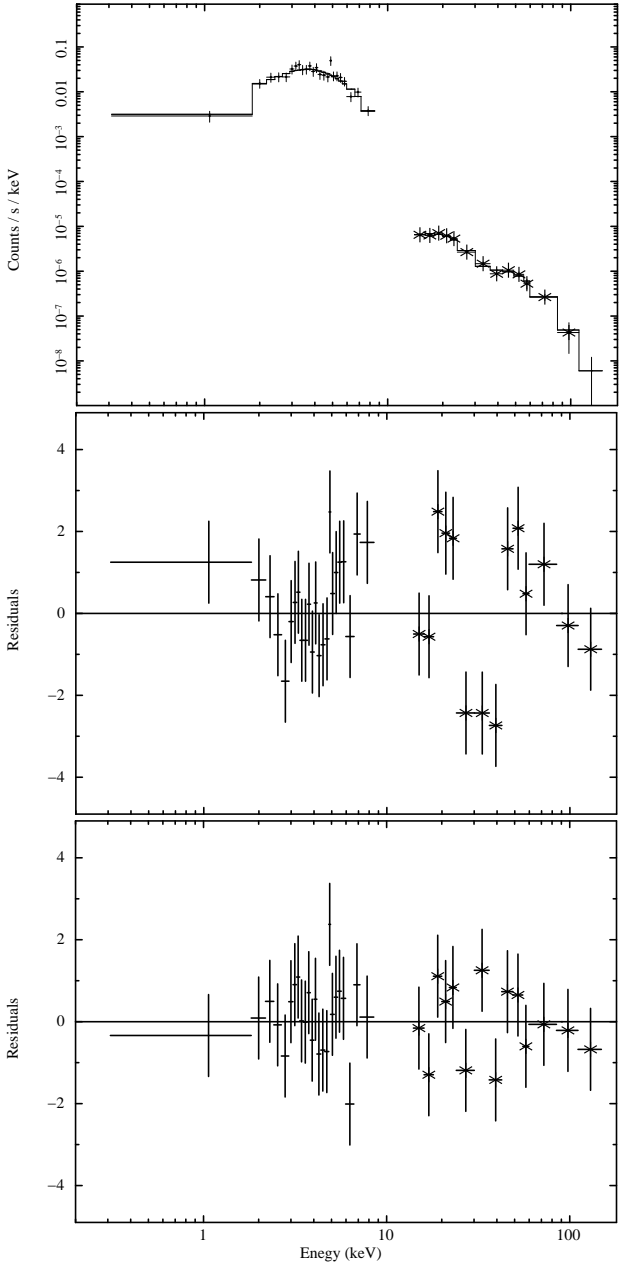

\begin{center}
\centerline{\includegraphics[width=5.4cm,angle=-90]{spettro_cyc_m.ps}}
\centerline{\includegraphics[width=5.4cm,angle=-90]{figura_residuals_nocyc_m.ps}}
\centerline{\includegraphics[width=5.9cm,angle=-90]{figura_residuals_cyc_m.ps}}
\caption[IGR~J19149+1036 XRT and BAT energy spectral distribution]{{\bf Top panel}:
IGR~J19149+1036 
XRT and BAT spectra and best fit {\tt phabs*cutoffpl} model. 
{\bf Middle panel}: Residuals in unit of standard deviations for the {\tt phabs*cutoffpl}
model.
{\bf Bottom panel}: Residuals in unit of standard deviations for the {\tt
phabs*cutoffpl*cyclabs} model.
                }
\label{spec} 
\end{center}
\end{figure}

In order to maximize the S/N ratio of the BAT data, we produced a spectrum selecting the data in
the phase interval 0.3-0.8 (see Figure~\ref{fold}, top panel ).
Then fitted the BAT spectrum and the XRT spectrum obtained from the entire XRT dataset using a
common spectral model, including a multiplicative factor that is frozen 
to unity for the XRT spectrum and is left free to vary  for the BAT  spectrum, to account for 
any inter-calibration uncertainty between the two telescopes and/or 
for different average source intensity between the  XRT and BAT observing times.   
The  combined  XRT-BAT  spectrum (Fig.~\ref{spec}, top panel)  was  first modeled  with  an
absorbed power law (model 1 {\tt cons*phabs*powerlaw}) with a $\chi^2$ of 74.7
with 34 degrees of freedom (d.o.f.). We also tried to model the broad band spectrum
including a cutoff in the model  (model 2 {\tt cons*phabs*cutoffpl}) obtaining a $\chi^2$  of 
69.9 (33 d.o.f.). 
Both models result indeed unacceptable, with residuals (Fig.~\ref{spec}, middle panel) that
show a broad range residual pattern both in the XRT and in the BAT data.
We added to both models an absorption cyclotron line ({\tt cyclabs}), obtaining a significant 
improvement in the $\chi^2$ (28.4, 30 d.o.f, with $\Delta\chi_0^2=41.5$) only for model 2 with a 
flat residual distribution (Fig.~\ref{spec}, bottom panel).
Table \ref{fit} reports the spectral results of the best fitting model.
To estimate the statistical significance of the presence of the cyclotron absorption feature 
we applied a Monte Carlo simulation on the BAT spectrum. The details of this
procedure can be found in \citet{dai11b}.
We find that the  probability of chance occurrence to find a $\Delta \chi^2$ value $\ge 41.5 $ is 
$\sim 2.7 \times10^{-6}$.

\begin{table}
\begin{tabular}{ r l l}
\hline
Parameter & Best fit value & Units    \\ \hline \hline
$\rm n_H$    & $3.9^{+0.3}_{-0.1} \times 10^{22}$ & $\rm cm^{-2}$\\
$\Gamma$  &$0.9^{+0.4}_{-0.4}$&          \\
$E_{cut}$ &$24^{+10}_{-5}$ & keV\\
$N$       &$ 1.8^{+1.6}_{-0.8}\times 10^{-3}$ &ph $\rm keV^{-1} cm^{-2} s^{-1}$ at 1 keV \\
$\rm C_{BAT}$&$1.1^{+0.5}_{-0.4}$&\\
$\rm E_{cyc}$& $31.2^{+1.9}_{-2.2}$& keV\\
$\rm D_{cyc}$& $0.9^{+0.4}_{-0.4}$& \\
$\rm W_{cyc}$& $8.4^{+4.2}_{-2.8}$& keV \\
$\rm F$ (0.3--10 keV)&$1.76\times 10^{-11}$& erg s$^{-1}$ cm$^{-2}$\\
$\rm F$ (15--150 keV)&$4.53\times 10^{-11}$& erg s$^{-1}$ cm$^{-2}$\\
$\chi^2$   &28.4  (30 d.o.f.) & \\ \hline
\end{tabular}
\caption{Best fit spectral parameters for the {\tt
phabs*cutoffpl*cyclabs} model.  $\rm C_{BAT}$ is the
constant factor to be multiplied to the model in order to match the XRT and BAT data.
We report {\bf observed} fluxes for the standard XRT (0.3--10 keV) and BAT (15--150 keV)
energy bands. \label{fit}}
\end{table}

\section{Conclusions}\label{conclusion}

We have analyzed the Swift-BAT and XRT data relevant to the HMXB IGR~J19149+1036. 
The source is close ($\sim$20 arcmin) to the much brighter source GRS 1915+105, and
an ad hoc imaging analysis was necessary to reveal it in the BAT all sky map, where it emerges
at a significance level of $\sim29$
($\sim32$) standard deviations in the 15--150 keV (15--60 keV) band after 
103 months of the Swift mission. 
The timing analysis on the long term BAT light curve 
unveils the presence of a periodic modulation in the hard X-ray emission 
with a period of P$_0=22.25\pm0.05$ days. We interpret this modulation
as the orbital period of the binary system. The folded light curve is characterized
by a single symmetric peak with a roughly sinusoidal shape that shows a minimum consistent with 
zero intensity, lasting $\sim30$ per cent of the orbital period.
Such a long minimum intensity phase interval is difficult to explain with 
a full eclipse of the compact object, unless we speculate that the system has both an extremely 
eccentric orbit and a high inclination angle. An alternative explanation is 
that the modulation is caused by the passage of the compact source through the 
stellar wind, whose density decreases as the source approaches the orbit apastron. 
However, no definite conclusion on the geometry of the binary system can be drawn, without knowing
the spectral type of the companion star. 

The broad band spectrum cannot be adequately described by the standard continua used for HMXB, because of
the presence of an absorption feature within the BAT energy range.
We obtain a good description of the data introducing a cyclotron line in the cutoff power law model.
The energy of the cutoff is $\sim 24$ keV and the line is centered at E$_{\rm cyc} \sim 31$ keV. 
The presence of absorption cyclotron lines has been revealed in several HMXB. 
They are thought to be originated near the magnetic poles of the
neutron star due to resonant scattering processes of the X-rays by electrons whose kinetic 
energies  are quantized in discrete Landau energy levels perpendicular to the B-field
\citep{Landau07}. 
Their energy centroid is related to the intensity of the magnetic field of the neutron star:
\begin{equation}
\rm B_{\rm obs} =10^{12} \times \frac{E_{\rm cyc}/(1 keV)}{11.6} Gauss\sim 2.7\times10^{12} Gauss
\end{equation}

If the cyclotron absorption takes place near the poles of the neutron star
the observed resonance energy shall be corrected by the effect of the gravitational
redshift: 
$\rm E_{\rm cyc}^{\rm obs}$ = $\rm E_{\rm cyc}(1 + \rm z)^{-1}$  with 

\begin{equation}
(1 + \rm z)^{-1} = \left(1-\frac{\rm 2GM_{\rm ns}}{\rm R_{\rm ns}c^2}\right)^{0.5}
\end{equation}

Using standard parameters for the neutron star ($M_{\rm ns}=1.4\,M_{\sun}$, $\rm R_{\rm ns} = 10^6$ cm),
the magnetic field at the neutron star surface would be B$_{\rm surf} \sim 3.6 \times 10^{12}$ 
Gauss. This result is fully consistent with the distribution of the magnetic field values derived for 
neutron stars in binary systems, that peaks in the range $1-4 \times 10^{12}$ G \citep{maki99}.
  
\section*{Acknowledgments}
This work has been supported by ASI grant I/011/07/0.

\bibliographystyle{aa}

{}

\end{document}